\documentclass[preprint,3p,twocolumn]{elsarticle}

\usepackage{graphicx,stmaryrd}
\usepackage{amsmath}
\usepackage{amssymb}
\usepackage{natbib}
\usepackage{hyperref}
\hypersetup{
    colorlinks = true,
    urlcolor   = blue,
    citecolor  = black,
}

\usepackage{xcolor}
\graphicspath{{figures/}} 
\usepackage{picture}

\usepackage{bm}
\usepackage{ulem}
\usepackage{cancel}
\usepackage[utf8]{inputenc} 
\usepackage[T1]{fontenc}    
\usepackage{textcomp}       
\usepackage{orcidlink}

\usepackage{tikz,xcolor,hyperref}

\newcommand\Rey{\mbox{Re}}
\newcommand\Deb{\mbox{De}}
\newcommand\Fr{\mbox{Fr}}
\newcommand{\frequency}{\omega}
\newcommand{\wavenumber}{k}
\newcommand{\texp}{\beta}
\newcommand{\sexp}{\alpha}

\newcommand{\meanflow}{U }
\newcommand{\flucts}{u }

\newcommand{\concentration}{\mathcal{B}}
\newcommand{\mB}{\mathcal{B}}

\def \TI  {T_{\rm I}}
\def \Tz  {T_{0}}

\def \urms {u_{\rm rms}}
\def \rr {\bm{r}}
\def \xx {\bm{x}}

\def \mF {\mathcal{F}}
\def \mC {\mathcal{C}}
\def \uu {\bm{u}}
\def \mS {\mathcal{S}}
\def \delt {\partial_{t}}
\def \fa   {f_{\alpha}}
\def \fb   {f_{\beta}}
\def \fg   {f_{\gamma}}
\def \deli {\partial_{i}}
\def \delj {\partial_{j}}
\def \delk {\partial_{k}}
\def \dela {\partial_{\alpha}}
\def \delb {\partial_{\beta}}
\def \delg {\partial_{\gamma}} 
\def \ui   {u_{i}}
\def \uj   {u_{j}}
\def \ua   {u_{\alpha}}
\def \ub   {u_{\beta}}
\def \ug   {u_{\gamma}}
\def \uk   {u_{\kappa}}
\def \ka   {k_{\alpha}}
\def \kb   {k_{\beta}}
\def \kg   {k_{\gamma}}

\def \pg   {p_{\gamma}}
\def \pa   {p_{\alpha}}

\def \pk   {p_{\kappa}}
\def \qg   {q_{\gamma}}

\def \dpdq {d^dpd^dq}

\def \Sij  {S_{ij}}
\def \Sab  {S_{\alpha\beta}}
\def \kk   {\bm{k}}
\def \pp   {\bm{p}}
\def \qq   {\bm{q}}
\def \mP   {\mathcal{P}}
\def \Ipq  {\int_{\pp,\qq}}
\def \ab   {\alpha\beta}
\def \muf  {\mu_{\rm f}}
\def \nuf  {\nu_{\rm f}}
\def \mup {\mu_{\rm p}}
\def \taup {\tau_{\rm p}}
\def \rhof {\rho_{\rm f}}
\def \mC  {{\Gamma}}
\def \Cij {C_{ij}}
\def \Cab {C_{\alpha\beta}}
\def \Cgg {C_{\gamma\gamma}}
\def \dij {\delta_{ij}}
\def \dab {\delta_{\alpha\beta}}
\def \mL  {\mathcal{L}}
\def \mU  {\mathcal{U}}
\def \mT  {\mathcal{T}}

\def \lap {\nabla^2}
\def \xhat {\hat{\bm{x}}}

\newcommand{\avg}[1]{\left\langle #1\right\rangle}
\newcommand{\eq}[1]{eq.~\ref{#1}}
\newcommand{\Fig}[1]{fig.~\ref{#1}}

\begin{document}

\begin{frontmatter}

\author[1]{Giulio Foggi Rota\,\orcidlink{0000-0002-4361-6521}} 
\author[1]{Rahul K. Singh\,\orcidlink{0000-0002-5399-1085}} 
\author[1,3]{Alessandro Chiarini\,\orcidlink{0000-0001-7746-2850}} 
\author[1]{Christian Amor\,\orcidlink{0000-0002-9710-7917}} 
\author[1]{Giovanni Soligo\,\orcidlink{0000-0002-0203-6934}} 
\author[2]{Dhrubaditya Mitra\,\orcidlink{0000-0003-4861-8152}}
\author[1]{Marco Edoardo Rosti\,\orcidlink{0000-0002-9004-2292}\corref{cor1}} 
\ead{marco.rosti@oist.jp}
\cortext[cor1]{Correspondence}
\affiliation[1]{organization={Complex Fluids and Flows Unit, Okinawa Institute of Science and Technology Graduate University}, addressline={1919-1 Tancha}, postcode={904-0495}, city={Onna-son}, country={Okinawa, Japan}}
\affiliation[2]{organization={Nordita, KTH Royal Institute of Technology and Stockholm University}, addressline={Hannes Alfvéns väg 12}, postcode={SE-106 91}, city={Stockholm}, country={Sweden}}
\affiliation[3]{organization={Currently at: Dipartimento di Scienze e Tecnologie Aerospaziali, Politecnico di Milano}, addressline={via La Masa 34}, postcode={20156}, city={Milano}, country={Italy}}

\title{The broken link between space and time in elastic turbulence}

\begin{abstract}
Elastic turbulence (ET), observed in flows of sufficiently elastic polymer solution at small inertia, 
is characterized by chaotic motions and power-law scaling of energy spectrum ($E$) 
in both wavenumber ($k$) and frequency ($\omega$): 
$E(k) \sim k^{-\alpha}$ 
and $E(\omega) \sim \omega^{-\beta}$. 
Experiments of ET have obtained a vast range of values for the exponent $\beta$. 
In inertial turbulence, Taylor's frozen-flow hypothesis implies $\alpha = \beta$, i.e., 
spatial and temporal scales are linearly related to each other. 
In contrast, from  high-resolution simulation in three different setups, 
a tri-periodic box, a channel, and a planar jet, we show that in ET $\alpha \approx 4$ 
while $\beta$ varies significantly. 
Our analysis shows that in general Taylor's  hypothesis does not hold in ET as there 
is no universal relation, linear or otherwise, between space and time. 
We thus clear the confusion of the different scaling exponents found in ET, 
and focus the attention of future research on understanding $\alpha$. 
{Our analysis also implies that waves-like dynamics 
with a linear dispersion relation (e.g., Alfv\'en waves) can not play a role in 
determining the scaling behavior of ET. 
The techniques introduced here can be useful for studying 
smooth chaotic flows in general, e.g., active turbulence.}
\end{abstract}

\end{frontmatter}

\section{Introduction}\label{sec:intro}
\begin{figure*}[t]
\centering
\includegraphics[width=\textwidth]{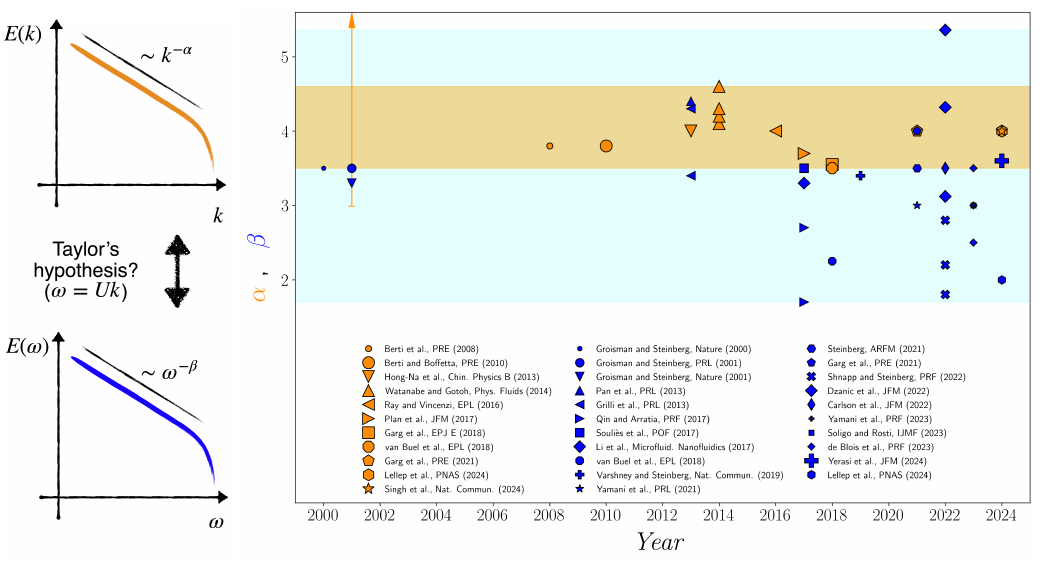}
\caption{\textbf{Power-law spectrum} of the turbulent kinetic energy measured in space and in time for different ET setups over the years. Orange symbols refer to spatial decay exponents $\sexp$ in the wavenumber domain ($\wavenumber^{-\sexp}$) and blue symbols to temporal decay exponents $\texp$ in the frequency domain ($\frequency^{-\texp}$). The only theoretical prediction is an inequality: $\sexp > 3$, denoted with an arrow. The ranges of variation of the exponents are shown as shaded regions, and highlight the sparsity of the measurements of $\texp$ compared to those of $\sexp$, which all lie in a narrow band centred around $4$. Notably, no trend for $\texp$ emerges even reordering the data in terms of the elasticity number (defined as the ratio of Deborah and Reynolds numbers).}
\label{fig:exponents}
\end{figure*}

\textit{Elastic turbulence} (ET) \cite{steinberg-2021} is the chaotic state attained by 
polymeric fluids in flow conditions where elasticity dominates over inertia. 
Elastic turbulence was first discovered in the early 2000s \cite{groisman-steinberg-2000}, and it has been 
investigated experimentally, numerically, and theoretically ever after.
Its footprint is a power-law decay of the energy spectrum  
$E(k) \sim \wavenumber^{-\sexp}$,
similar to what is observed in \textit{inertial turbulence} 
(IT) \cite{kolmogorov-1941, frisch-1995}, although the actual exponent of the 
power-law is different.
Yet, we note that most experimental results in ET are time series of the 
fluctuating velocity at one point in space, yielding a frequency-dependent 
energy spectrum with power-law scaling 
$E(\omega) \sim \omega^{-\texp}$. 
On one hand, both experimental and numerical results~\cite{groisman-steinberg-2000,groisman-steinberg-2001-1, groisman-steinberg-2001-2,pan-etal-2013,grilli-vazquezquesada-ellero-2013,quin-arratia-2017,soulies-etal-2017, varshney-steinberg-2019, yamani-etal-2021,steinberg-2021, shnapp-steinberg-2022, dzanic-from-sauret-2022, carlson-etal-2022, yamani-etal-2023, soligo-rosti-2023, deblois-haward-shen-2023, yerasi-etal-2024} 
report a vast range of values for the exponent $\texp$, as shown in 
figure~\ref{fig:exponents}.
On the other hand, numerical and theoretical investigations~\cite{balkovsky-fouxon-lebedev-2001, berti-etal-2008,berti-boffetta-2010, hongna-etal-2013, watanabe-gotoh-2014,ray-vincenzi-2016,li-etal-2017, plan-etal-2017, garg-etal-2018-2, vanbuel-schaaf-stark-2018, garg-calzavarini-berti-2021, lellep-linkmann-morozov-2024, singh-etal-2024} identify a narrower range for $\sexp$, with recent 
works converging towards a power-law decay in the wavenumbers with 
exponent $\sexp \approx 4$ common to multiple flow 
configurations \cite{berti-boffetta-2010, lellep-linkmann-morozov-2024, 
singh-etal-2024, garg-rosti-2025} (with few exceptions in two dimensions, explained in \ref{app:pope}).

Simulations provide access to the whole flow field, allowing for a direct 
computation of the spatial as well as the temporal energy spectra, and thereby 
a measurement of both the exponents $\sexp$ and $\texp$.
In experiments, it is instead common to work with temporal measurements, 
recovering the temporal exponent $\texp$.  
The two exponents are typically related applying \textit{Taylor's frozen 
turbulence hypothesis},
proposed by G. I. Taylor in 1938 \cite{taylor-1938} to relate the spatial and 
temporal statistics of grid turbulence in Newtonian flows. 
He reasoned that, if the intensity of the turbulent velocity fluctuations 
$\flucts$ is small compared to the mean flow speed $\meanflow$, then the 
temporal response of the flow at a fixed point in space can be seen as the 
result of an unchanging spatial pattern convecting uniformly past that point 
at velocity $\meanflow$.
In other words, if Taylor's hypothesis holds, there is a linear relationship 
between space (or the wavenumber $\wavenumber$) and time (or the frequency 
$\frequency$) mediated by the mean flow speed $\meanflow$, such that 
$\frequency=\meanflow\wavenumber$, leading to $\alpha = \beta$.
Taylor's hypothesis can be extended to flows without a mean velocity too: 
indeed, an eddy of size $1/k$ is swept by a large scale eddy of size $1/k_0$ 
over a time $1/(k u_{k_0})$, thus leading again to 
$\omega \sim k u_{k_0} \sim k$, and $\sexp = \texp$ 
\cite{frisch-1995, mitra-pandit-2003}. 
Note that, in most cases the scaling range of the spatial as well as 
temporal energy spectra is barely a decade, thus the scaling exponents cannot 
be determined with high accuracy.
Nevertheless, the large scatter shown in figure~\ref{fig:exponents} 
raises questions on the applicability of Taylor's hypothesis in ET, and cannot be attributed to a limited scaling range only. 
This may be attributed to either of the following two reasons: 
first, the hypothesis holds in ET but $\sexp$ and $\texp$, even though not 
determined with high accuracy, are \textit{not universal} and depend on how 
the flow is driven;
second, the hypothesis cannot be applied to ET and there is no universal 
relation, linear or otherwise, between space and time. 
Here we demonstrate, by analyzing direct numerical simulations of different 
flows, that the latter is true: Taylor's hypothesis does not hold in ET.

\section{Model}\label{sec:mod}
\label{sec:model}
Let us start by writing down the governing equations for a dilute polymeric fluid~\citep{oldroyd-1950, bird-1987}, i.e., the Navier--Stokes equations 
coupled to the evolution equation of the averaged polymers' configuration, described by the Oldroyd-B model: 
\begin{subequations}
  \begin{align}
    \rhof\left( \delt \ui + \uj \delj \ui \right) &=
    -\deli p +
    \delj \left( 2 \muf\Sij  + \frac{\mup}{\taup} \Cij\right), \
     \label{eq:NN}  \\ 
    \delt\Cij + \uk\delk\Cij &= C_{ik}\delk \uj + C_{k j}\deli\uk \notag\\
    - \frac{1}{\taup}\left(\Cij - \dij\right)  \/.\label{Conf} 
  \end{align}
  \label{eq:governing}
\end{subequations}
Here, $\uu$ is the incompressible velocity field, with $\delj \uj = 0$, 
$p$ is the pressure, $\mathbf{S}$ is the rate-of-strain tensor with components 
$\Sij \equiv (\deli\uj + \delj\ui)$, $\muf$ and $\rhof$ are the dynamic 
viscosity and the density of the solvent fluid with 
$\nuf = \muf/\rhof$ being its kinematic viscosity, 
$\mup$ and $\taup$ are the polymer viscosity
and relaxation time, and $\mathbf{C}$ (with components $\Cij$) 
is the polymer conformation tensor whose trace $C_{kk}$ is the total 
end-to-end averaged squared length of the polymer.
We use Einstein's notation of repeated indices being summed.
Three dimensionless numbers govern the behaviour of the 
equations \ref{eq:governing}: 
the Reynolds number $\Rey = \mU\mL/\nuf$, 
the Deborah number $\Deb = \mU\taup/\mL$, and the 
solvent-to-total viscosity ratio $\mB = \muf/(\mup+\muf)$,
where $\mU$ and $\mL$ are the characteristic velocity and length scales of the 
problem.
We choose $\mB$ close to unity, as we simulate dilute polymer solutions.
To study ET, we impose the Reynolds number to be small and the Deborah number 
to be sufficiently large. 
We consider three different flow configurations, with increasing 
complexity: 
a triperiodic box with an external stirring, a pressure-driven channel flow, 
and a planar jet. 
All of them were individually investigated in our former 
works~\cite{singh-etal-2024, foggirota-etal-2024-3, soligo-rosti-2023}. 

\section{Setup and Methods}\label{sec:snm}

\subsection{Box}

Our box simulations are performed in a cubic domain of edge-length $\mL = 2\pi$, uniformly discretised on $512$ grid points along each direction.
Periodicity is enforced at each boundary, and the flow develops under an Arnold-Beltrami-Childress (ABC) forcing \cite{dombre-etal-1986} with unitary amplitudes. 
To confirm the independence of our results from the forcing adopted, we replicated them with the one proposed by Eswaran and Pope \cite{eswaran-pope-1988}, observing no significant variation in the spatial and temporal spectra of the turbulent kinetic energy (see \ref{app:pope}).
The ET flow is characterised by a Taylor-scale Reynolds number $Re \approx 40$, a Deborah number $De=9$, and a viscosity ratio $\concentration=0.9$  \citep[as in our former study][]{singh-etal-2024}. 
To compute $Re$ and $De$ we employ the RMS of the velocity fluctuations $\mU$, the edge-length $\mL$, the total viscosity and the relaxation time of the polymer.
The IT flow introduced for comparison is characterised by $Re \approx 145$.
In both the ET and IT cases, the energy spectra are computed performing an isotropic average across the whole domain and over 40 successive snapshots {homogeneously sampled in $490\mT$, with $\mT = \mL/\mU$}; the spectra of all three velocity components are summed and integrated over a shell in Fourier space. 
The temporal fluctuations of the velocity used to attain the temporal spectra and the structure functions, instead, are sampled {over the same time} with an array of $128^2$ probes, evenly spaced in a plane crossing the centre of the domain.
{In this setup, as in all the following, we aimed at ensuring sufficient sampling points for averaging temporal statistics and sufficient proximity between the sampling points to resolve at least the largest scale over which fluctuations remain correlated (with acceptable computational burden). We thus homogeneously distributed the probes every 4 grid points in the selected plane.}

\subsection{Channel}
Our channel simulations are performed in a rectangular domain of size $4\pi \mL \times 2 \mL \times 2\pi \mL$ along the streamwise, wall-normal and spanwise directions, uniformly discretised on $512 \times 1024 \times 256$ grid points in the ET case, and in a domain of size $5\mL \times 2\mL \times 2\mL$ uniformly discretised on $1000 \times 400 \times 400$ grid points in the IT case. $\mL$ denotes the half-height of the channel.
{The IT channel simulations are consistent with the minimal flow unit \cite{jimenez-moin-1991} needed to develop and sustain inertial turbulence, resolving all flow structures on a staggered cubic grid with the first node at $z^+=0.9$. Instead, in ET, flow structures are different with large coherent motions spanning the centre of the channel \cite{buza-etal-2022} and small-scale perturbations at the walls \cite{beneitez-page-kerswell-2023}. We thus employ a larger domain and a finer grid in the wall-normal direction compared to ET to appropriately resolve such features \cite{foggirota-etal-2024-3}.}
No-slip and no-penetration boundary conditions are imposed at the top and bottom walls, while periodicity is enforced along the streamwise and spanwise directions. 
The flow develops under a uniform streamwise pressure gradient, adjusted in time to ensure a constant flow rate.
The ET flow is characterised by a bulk Reynolds number $Re=5$, a Deborah number $De=50$, and a viscosity ratio $\concentration=0.9$
\citep[as in our former study][]{foggirota-etal-2024-3}. 
To compute $Re$ and $De$ we employ the bulk velocity $\mU$, the half-channel height $\mL$, the total viscosity and the relaxation time of the polymer.
The IT flow introduced for comparison is characterised by $Re = 5000$.
In both the ET and IT cases, the energy spectra are computed at the centre-plane along the streamwise direction, averaged in the spanwise direction and over 220 successive snapshots  {homogeneously sampled in $100\mT$}, summing the spectra of all three velocity components. 
The temporal fluctuations of the velocity used to attain the temporal spectra and the structure functions are also sampled at the centre-plane {over the same time}, with an array of $64 \times 32$ probes in the ET case and $250 \times 100$ probes in the IT case, evenly spaced along the streamwise and spanwise directions, respectively. 

\subsection{Jet}

Our jet simulations are performed in a rectangular domain of size $160 \mL \times 240 \mL \times 53.3 \mL$ along the streamwise, jet-normal and spanwise directions, uniformly discretised on $1536 \times 2304 \times 512$ grid points in the ET case, and in a domain of size $160 \mL \times 160 \mL \times 40 \mL$, uniformly discretised on $3280 \times 3280 \times 820$ grid points in the IT case.
The flow is injected from a slit with half-width $\mL$ on the left side of the domain at a constant velocity distributed in a plug; the polymers are at rest when injected in the domain. 
No-slip and no-penetration boundary conditions are enforced at the left wall, exception made for the inlet slit. Free-slip boundary conditions are imposed at the top and bottom boundaries, and a non-reflective outflow is implemented at the right boundary \cite{orlanski-1976}. Periodicity is assumed along the span-wise direction.
The ET flow is characterised by an inlet Reynolds number $Re=20$, a Deborah number $De=100$, and a viscosity ratio $\concentration=0.98$  \citep[as in our former study][]{soligo-rosti-2023}. 
To compute $Re$ and $De$ we employ the velocity at the inlet $\mU$, the half-height of the slit through which the fluid is injected $\mL$, the total viscosity and the relaxation time of the polymer.
The IT flow introduced for comparison is characterised by $Re = 4500$.
In both the ET and IT cases, the energy spectra are computed at the centre-plane along the spanwise direction at a streamwise distance from the inlet of $56\mL$ such to ensure fully developed turbulence, averaged over 140 successive snapshots {homogeneously sampled in $340\mT$}, summing the spectra of all three velocity components.
The temporal fluctuations of the velocity used to attain the temporal spectra and the structure functions are also sampled at the same location {and over the same time}, with a line of $128$ probes in the ET case and $205$ probes in the IT case, evenly spaced along the spanwise direction.

\subsection{Numerical Method}
The direct numerical simulation of a viscoelastic fluid flow in ET poses
significant challenges.
The high-accuracy numerical schemes required for a correct integration of the governing equations and the computational power needed to ensure an adequate resolution make this kind of simulations a remarkable endeavor \cite{doubief-terrapon-hof-2023}.  
In this study, the problem is accurately and efficiently tackled by means of our well validated solver \href{https://www.oist.jp/research/research-units/cffu/code}{Fujin} \cite{rosti-perlekar-mitra-2023, abdelgawad-cannon-rosti-2023,soligo-rosti-2023}.
We adopt a staggered uniform Cartesian grid and discretize equations \ref{eq:governing} in space according to a second-order central finite-difference scheme; a second-order Adams–Bashforth scheme is chosen for time integration, coupled with a fractional step method \cite{kim-moin-1985}. 
The transport equation for the polymer conformation tensor $\mathbf{C}$ requires additional care: we tackle it in a matrix-logarithm formulation \cite{fattal-kupferman-2005,devita-etal-2018} to overcome the notorious high Deborah numerical instability and resort to a high-order weighted essentially non-oscillatory (WENO) scheme \cite{sugiyama-etal-2011} to treat the upper-convected derivative constituting the left-hand side of equation~\ref{eq:governing}b.
In this way we are able to avoid the introduction of a stabilising stress-diffusion term in equation~\ref{eq:governing}b, which therefore does not require us to specify any boundary condition for the polymer \cite{beneitez-etal-2024}.

\section{Results}\label{sec:res}
\begin{figure*}[htb]
\centering
\includegraphics[width=.90\textwidth]{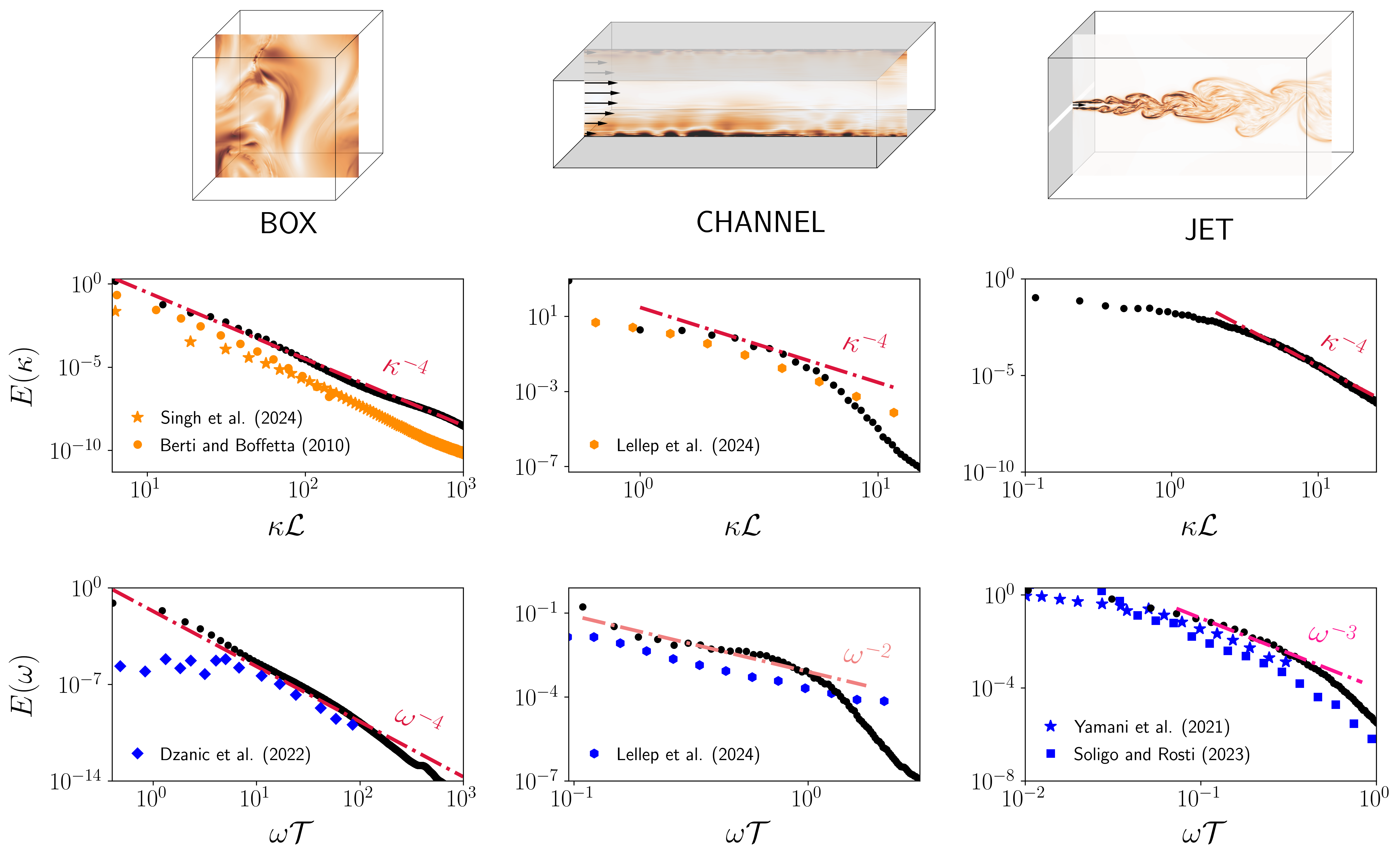}
\caption{\textbf{Turbulent kinetic energy spectra} of elastic turbulence, for box (left), channel (middle) and jet (right),  from left to right. 
Top row: visualizations of the vorticity magnitude, with darker colors denoting higher values. 
Middle row: spatial energy spectra, normalized with the mean-square of the velocity fluctuations. 
Bottom row: temporal energy spectra, normalized with the mean square of the velocity fluctuations.
\textit{Our data is marked with black dots}, while data from literature is highlighted with the same symbols of figure~\ref{fig:exponents} and colors consistent with their spatial/temporal nature. 
A slight shift downward is {applied only to literature data} for ease of visualization. Power-law scalings are reported with dashed-dotted lines.}
\label{fig:spectra}
\end{figure*}

The spatial spectra of the turbulent kinetic energy for 
the instances of a periodic box, a channel, and a jet  exhibit a power-law decay of  
$E(\wavenumber) \sim \wavenumber^{-\alpha}$ with $\alpha=4$, 
in agreement with previous studies, reported in figure \ref{fig:spectra}. 
However, the scaling range for the channel and the jet is narrower 
(less than a decade) compared to that of the 
box (almost two decades).
This result remains unchanged even if \textit{(i)} 
we change the model of polymers from Oldroyd-B to 
FENE-P~\cite{singh-etal-2024, foggirota-etal-2024-3} 
or PPT \cite{lellep-linkmann-morozov-2024}, 
\textit{(ii)} $\Rey$ is varied within the range $1500$ to $0.5$ 
in the channel~\citep{foggirota-etal-2024-3}, 
and \textit{(iii)} even holds in two dimensions~\citep{berti-boffetta-2010}. 
Yet, the temporal spectra exhibits different 
$E(\frequency) \sim \frequency^{-\beta}$ scaling behaviors, 
with $\beta=4$ in the box, $\beta=2$ in the channel, and $\beta=3$ in the jet. 
Importantly, for a fixed setup, the scaling of the temporal spectra is robust, 
e.g.~$\omega^{-3}$ is observed in jets 
for a large range of $\Rey$~\cite{yamani-etal-2021, soligo-rosti-2023}, from 
$800$ to $20$. Note that previous investigations \cite{dzanic-from-sauret-2022} have shown that the slope $\texp$ varies with the Schmidt number $Sc$, quantifying polymeric diffusion.
Here, however, $Sc \rightarrow +\infty $ as we do not include any explicit diffusion in equation~\ref{eq:governing}b apart from that implicit in the numerics, so our results do not depend from $Sc$.
We thus conclude that, in general, Taylor's hypothesis does not hold in ET. 
In the following, we focus on the channel and the box as prototypical flows 
with and without a mean velocity.

\subsection{A phenomenological approach to the breakdown of Taylor's hypothesis}
Let us now revisit Taylor's hypothesis in inertial turbulence --
high Reynolds number and no polymers -- by using phenomenological
arguments ``\`a la Kolmogorov''.
By Kolmogorov's phenomenology, the characteristic velocity 
$u(r)$ at scale $r \ll \mathcal{L}$ is
given by $u/\mU \sim (r/\mL)^h$ with $h=1/3$. 
Here, $\mU$ is the characteristic velocity at the large scale $\mL$.
We start by considering equation~\ref{eq:NN} without the term containing the 
polymeric stress.
The remaining terms yield three distinct characteristic time-scales: 
advection by velocities at large ($\mL$) and small ($r$) scales give the 
estimates $r/\mU$, and $r/u$ whereas viscous dissipation gives $r^2/\nuf$.
These three timescales can be rewritten as
$(r/\mL) \mT $, $(r/\mL)^{1-h} \mT $, and $(r/\mL)^2\Rey \mT$,
recalling $\mT = \mL/\mU$.
We take the limit $\Rey \to \infty$ and $(r/\mL)\to 0$, thus the
fastest timescale is $(r/\mL) \mT $,
i.e.,
$\tau(r) \sim (r/\mL) \mT$.
This linear relationship,
$\tau(r) \sim r^{z}$ with $z=1$,
between time and space is essentially Taylor's hypothesis 
(it implies $ \frequency \sim \wavenumber$), and
the exponent $z$ is called ``dynamic exponent''.
We mention in passing that this relationship is true only for Eulerian 
velocities; in Lagrangian~\citep{kaneda1999taylor} or
quasi-Lagrangian~\citep{belinicher1987scale, mitra2005dynamics} frames
the advection by the large scale velocity can be removed, and within the
Kolmogorov model of simple scaling $\tau(r) \sim r^{1-h}$ emerges.

Extending these arguments to ET, we now find four timescales:
$(r/\mL) \mT $, $(r/\mL)^{1-h} \mT$, $(r/\mL)^2 \Rey \mT$, and
$(r/\mL)^2\Deb\Rey\mT$.
In ET, the velocity field is smooth and $u(r) \sim r^h$ with $h=1$. For 
a fixed (small) value of $\Rey$ and $\Deb \gg 1$,
in the limit $(r/\mL) \to 0$,  the fastest timescale is thus 
$\tau(r) \sim r^2$, which emerges from the viscous term. 
In practice, however, the presence of a constant ($r^0$), which is likely 
case-dependent, causes no dynamic exponent to emerge: in ET the linear 
relation between time and space is broken and Taylor's hypothesis does 
not hold. 

\begin{figure*}[htb]
\centering
\includegraphics[width=\textwidth]{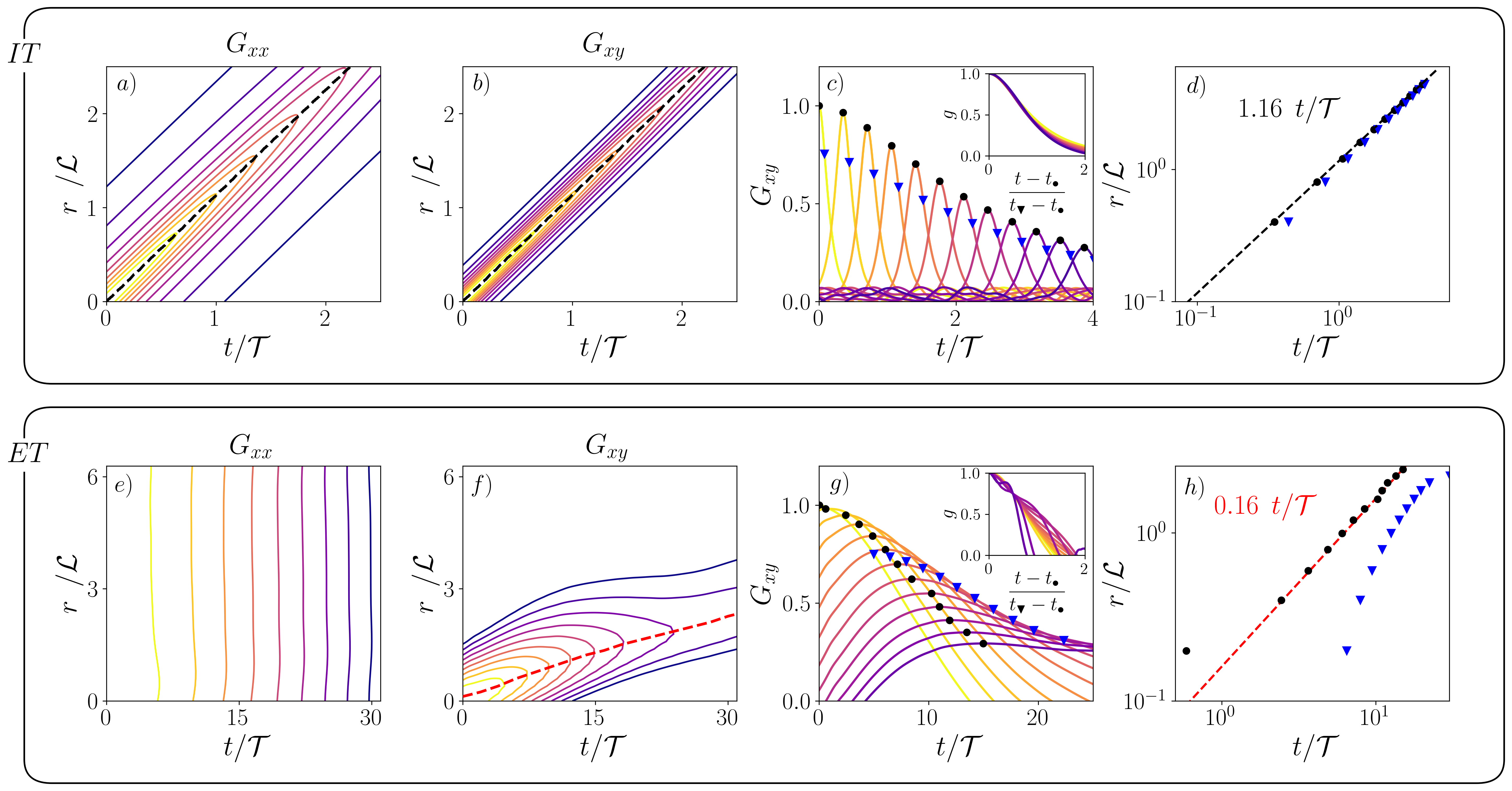}
\caption{\textbf{Space-time relationship in the IT (top row) and ET (bottom row) channel flows.} Contour plots (\textit{a,e}) of the streamwise
 autocorrelation of the streamwise velocity component, $G_{xx}(r,t)$,
 see equation~\ref{eq:Cxx}, and (\textit{b,f}) streamwise autocorrelations of the
 spanwise velocity component, $G_{xy}(r,t)$, as a function of $r/\mL$ and
 $t/\mT$.
 The autocorrelations are computed at the center-plane.
 The contour lines are evenly spaced between $0.1$ (blue) and $0.9$ (yellow)
 times the maximum.
We also plot (\textit{c,g}) $G_{xy}$ as a function of $t$ for several
different values of $r$.
Black dots demarcate the time of the maximum of $G_{xy}$, $t_\bullet$.
Blue triangles mark the time, $t_{\blacktriangledown}$, at which $G_{xy}$ falls below a
threshold value of $0.75$.
The insets show the attempted collapse following the ansatz in equation~\ref{eq:Gcol},
assuming a linear relationship between space and time.
We indeed obtain data collapse in IT, but not in ET.
Finally, we plot (\textit{d,h}) the two time-scales extracted from
$G_{xy}$ for different length scales. For IT they exhibit the same linear
behavior, but for ET different time scales follow different behaviors.}
\label{fig:correla}
\end{figure*}

\subsection{Space-time relationship in the channel}
To develop a clear understanding of the mechanism that leads to the breakdown
of Taylor's hypothesis in ET, we revisit the channel flow case comparing and
contrasting ET with IT at a Reynolds number $\Rey = 2800$.
We choose a coordinate system such that $x$ is along the streamwise direction,
$y$ is along the spanwise direction, and $z$ is along the wall-normal direction.
The domain is periodic in $x$ and $y$, bounded by walls in $z$.
A common way of testing the validity of Taylor's hypothesis in channel flows is
by probing the space-time correlation functions of the velocity fluctuations $\uu$ 
\begin{subequations}
  \begin{align}
  \mC_{xx}(r,t) = \avg{u_x(\xx+\xhat r,t)u_x(\xx,0)}\/, \notag\\
  \mC_{xy}(r,t) = \avg{u_y(\xx+\xhat r,t)u_y(\xx,0)}\/;\\
  G_{xx}(r,t) = \frac{\mC_{xx}(r,t)}{\mC_{xx}(0,0)}\/, \notag\\
  G_{xy}(r,t) = \frac{\mC_{xy}(r,t)}{\mC_{xy}(0,0)}\/;
  \label{eq:Cxx}
  \end{align}
\end{subequations}
where the first subscript indicates the direction along which the correlation 
is evaluated, and the second the component of velocity considered.
Here $\xhat$ is the unit vector along the streamwise direction, 
and averages $\langle \cdot \rangle$ are done in time and along the 
homogeneous directions, leaving only the dependence on the space and time separations, $r$ and $t$.
The correlations at the center-plane of the channel relate to the space and 
time energy spectra plotted in figure~\ref{fig:spectra} as
\begin{equation}
\begin{split}
  E(k) = \frac{1}{2}\int \sum_{i}\mC_{xi}(r,0)\exp(\mathrm{i}kr) dr, \\
  E(\omega) = \frac{1}{2} \int \sum_{i}\mC_{xi}(0,t)\exp(\mathrm{i}\omega t) dt,
\end{split}
\end{equation}
respectively. 

\begin{figure*}[htb]
\centering
\includegraphics[width=\textwidth]{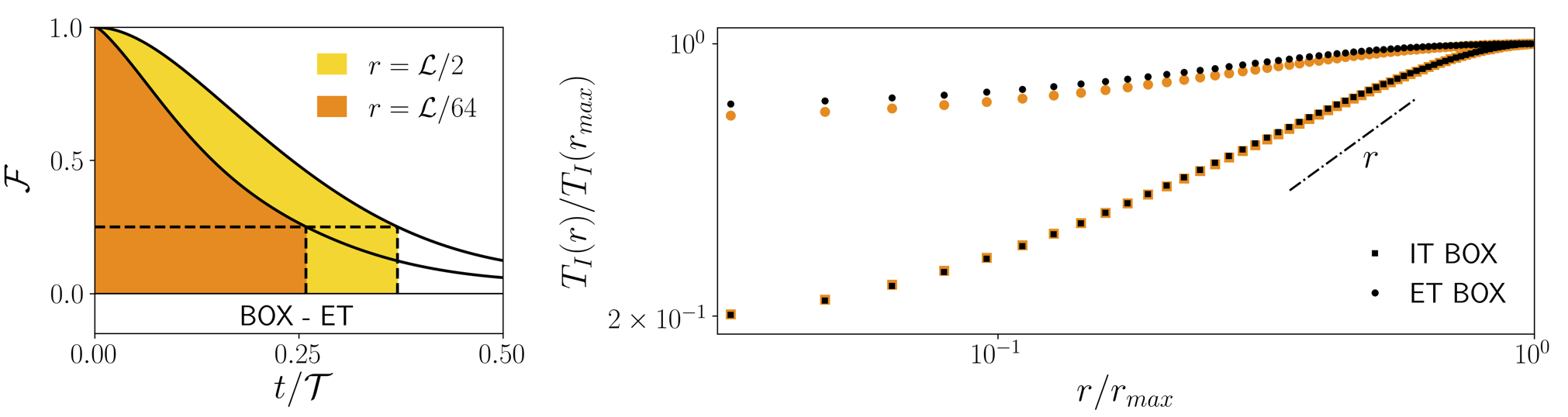}
\caption{\textbf{Dynamic scaling in the box.}
  (left) A representative plot of $\mF(r,t)/\mF(r,0)$ for two values of $r$
  for ET in the box. The dashed horizontal line is at the value of
  ordinate equal to $0.25$. The shaded area is the integral
  scale $\TI$ defined in equation~\ref{eq:TI} for the case where the threshold
  is $0.25$.
  (right) The characteristic timescale $\TI(r)$ as a function of $r$
  for IT (squares) and ET (circles); blue symbols are
  for a threshold of $0.25$ and black symbols are for the zero--crossing. 
  Note that, the black squares of IT overlap the blue ones. 
  A dashed line with unitary slope is shown to guide the eye. 
}
\label{fig:structure}
\end{figure*}

In figure~\ref{fig:correla}a we show the contour plots of 
$G_{xx}(r,t)$ as a function of $r$ and $t$. 
If Taylor's hypothesis holds, then the slope of the line connecting the 
spatial maxima of the contours 
is proportional to 
the average advective velocity with 
which fluctuations are transported downstream \cite{quadrio-luchini-2003}.
At the centerline of the IT channel flow, such velocity is,
approximately,
$1.16 \enspace \mU$ \cite{kim-moin-moser-1987}, 
where $\mU$ for a channel is the mean velocity of the flow.
The picture is dramatically different in ET, see figure~\ref{fig:correla}e. 
Here, the contours are parallel to the ordinate, so no dynamic scaling is 
observed.
Now, let us compare IT and ET for the other correlation function,
$G_{xy}(r,t)$.
In IT, the contours of figure~\ref{fig:correla}b make it clear that a dynamic
exponent $z=1$ holds.
Yet, the situation in ET is again quite different.
The peaks of the contours in figure~\ref{fig:correla}f  seem to follow an 
approximately linear space-time relation, 
but the remaining regions of the contours do not; 
suggesting that a unique relation between space and time may not exist. 
To investigate this further, we plot $G_{xy}$ as a
function of $t$ for several different values of $r$ in IT and ET in 
figure~\ref{fig:correla}c,g.
In IT, the plots suggest the following data collapse:
\begin{equation}
\begin{split}
  G_{xy}(r,t) = g\left[\frac{t-t_\bullet(r)}{t_{\blacktriangledown}(r) -t_\bullet(r)}\right]
  \quad\text{with} \\
  t_\bullet(r) \propto r \quad\text{and}\quad t_{\blacktriangledown}(r)\propto r,
  \label{eq:Gcol}
\end{split}
\end{equation}
where $t_\bullet(r)$ is the time at which $G_{xy}$ peaks and $t_{\blacktriangledown}(r)$ is 
any other timescale one may choose. 
In the following, we set $t_{\blacktriangledown}(r)$ to be the time at which 
$G_{xy}$ falls below the threshold value of $0.75$.
We check in the inset of figure~\ref{fig:correla}c that this collapse indeed
holds.
In figure~\ref{fig:correla}d we plot both $t_\bullet(r)$ and $t_{\blacktriangledown}(r)$
on the abscissa, and the corresponding $r$ values on the ordinate.
In both cases we obtain a clean linear behaviour.
This conclusively proves that in the IT channel flow the dynamic exponent is
$z=1$.
We now try the same approach for ET, plotting the values of
$t_\bullet(r)$ from figure~\ref{fig:correla}g in figure~\ref{fig:correla}h.
They do show a linear trend, but we do not find a collapse for the complete
correlation function -- see the inset of figure~\ref{fig:correla}g. 
Next, similar to IT, we also extract 
$t_{\blacktriangledown}(r)$ from $G_{xy}$ in figure~\ref{fig:correla}g and 
plot its values in figure~\ref{fig:correla}h, with $r$ on the ordinate.
We do not find any significant scaling, again demonstrating that there is no 
dynamic scaling in ET -- the link between space and time is broken.

\subsection{Space-time relationship in the box}
To develop a clearer understanding of the relationship between space and time
in ET, we now consider the simple case of a triperiodic box, where the
problem is homogeneous and isotropic.
Following earlier attempts~\citep{l1997temporal, mitra2004varieties,
  ray2011dynamic, biferale-calzavarini-toschi-2011} to understand the
space-time nature of IT, we define,
\begin{subequations}
  \begin{align}
    \mC(r,t) &= \avg{\ui(\xx+\rr,0)\ui(\xx,t)}\/,\\
    \delta_{\rr} \ui(t) &=  \ui(\xx+\rr,t)-\ui(\xx,t)\/,\\
    \label{eq:STF}
    \mF(r,t) &= \avg{\delta_{\rr}\ui(0)\delta_{\rr}\ui(t)} = 2\mC(0,t) - 2\mC(r,t) \/.
  \end{align}
\end{subequations}
The functions $\mC(r,t)$ and $\mF(r,t)$ are the space and time dependent
second order correlation function and structure function, respectively.  
They do not depend on $\xx$ because of homogeneity, and they only depend on the
magnitude of $\rr$, and  not on its direction, because of isotropy.
The two energy spectra shown before are related to the correlation function by
\begin{equation}
\begin{split}
    E(k) = \frac{1}{2} \int \mC(r,0)\exp(\mathrm{i}kr)dr, \\
    E(\omega) = \frac{1}{2} \int \mC(0,t)\exp(\mathrm{i}\omega t)dt.
\end{split}
\end{equation}
From the time-dependent correlation function, we can define a
scale-dependent integral time scale
\begin{equation}
  \TI(r) \equiv \frac{1}{\mF(r,0)}\int_0^{\Tz} \mF(r,t)dt\/,
    \label{eq:TI}
\end{equation}
where $\Tz$ is the time it takes to cross a threshold.
We use two values for this threshold, $0.25$ and $0$. 
Taylor's hypothesis would imply $\TI(r) \sim r^z$ with $z=1$. 
In figure~\ref{fig:structure} we plot $\TI(r)$ as a function of $r$
in a log-log scale for both IT and ET.
In the case of IT, we obtain a small range over which $z=1$.
Similar results have been reported
before~\citep{kaneda-ishihara-gotoh-1999,biferale-calzavarini-toschi-2011}. 
For ET, instead, we observe that it is difficult to find any significant
scaling range, with the function being almost a constant ($z=0$), 
especially at large $r$.
Again, Taylor's hypothesis does not hold; the link between space and time
is broken in ET, 
notwithstanding that both spatial and temporal spectra have 
$\sexp = \texp = 4$ in the box.

It is possible to develop an approximate understanding of this result
in the extreme case of $\Rey\to 0$.
In this limit, equation~\ref{eq:NN} reduces to the linear, forced, Stokes 
equation, which can be solved by Fourier transform.
Substituting the solution back in equation~\ref{Conf}, we obtain
a nonlinear evolution equation for the tensor $\Cij$.
A straightforward scaling analysis of this equation, with the
assumption that the scaling exponent for the velocity is $h=1$ in ET,
yields two possible values: $z=0$ and $z=2$, see appendix~\ref{app:theory}.
At large $r$, $z=0$ dominates over $z=2$, which is indeed what
we observe. 

\section{Discussions}\label{sec:end}
Several comments are now in order. 
First, note that the key feature that makes ET different from IT
is that the advective nonlinearity is sub-dominant in ET.
We verified that it is reasonable to completely remove the
nonlinear advective term from the momentum equation as an approximation,
by re-running our ET simulations in the
box and the channel with the nonlinear advection turned off
(see appendix~\ref{app:nlAdv}); the energy spectra of the velocity
remained practically unchanged.
It is thus reasonable to ignore the nonlinear advection in the
momentum equation to develop a theoretical description of ET.

Second, multiple investigations in IT have discussed the limitations of
Taylor's hypothesis and
possible improvements~\cite{lin-1953,lumley-1965,piomelli-balint-wallace-1989,
  cenedese-romano-difelice-1991, lee-lele-moin-1992, dahm-southerland-1997,
  belmonte-martin-goldburg-2000,
  ganapathisubramani-lakshminarasimhan-clemens-2007,
  moin-2009, davoust-jacquin-2011, hutchins-marusic-2007, dennis-nickels-2008,
  geng-etal-2015,cheng-etal-2017,roy-miller-gunaratne-2021}.
It has for example been noticed that, in grid turbulence, Taylor's hypothesis
does not accurately represent the dilatational part of the
flow~\citep{lee-lele-moin-1992}, while it also breaks down
in flows with high shear~\citep{lin-1953}.
Issues similar to those encountered when applying Taylor's hypothesis to IT
in the presence of coherent structures have been reported also in
ET~\citep{burghelea-segre-steinberg-2005},
and caution is recommended when investigating higher-order moments based on
temporal sampling only~\citep{garg-calzavarini-berti-2021}.
{However, here we show that the use of Taylor's 
hypothesis in ET is fundamentally flawed, unlike in IT.}

Third, note that in IT, in a \textit{Lagrangian or quasi-Lagrangian} frame,
the dynamical scaling is broken -- there is no unique dynamic exponent and 
dynamical multiscaling holds~\citep{l1997temporal, mitra2004varieties,
  ray2011dynamic, de2024uncovering} 
due to the fundamental multifractal nature of IT.
This is not relevant to the breakdown of Taylor's hypothesis, as we have
discussed here.

Fourth, the existence of a  dynamic exponent $z=1$ may also mean existence of
Alfv\'en-like waves in ET, as suggested in Ref.~\citep{fouxon-lebedev-2003}.
To investigate the possibility of elastic waves (with
the same dispersion relation as Alfv\'en waves), Varshney
and Steinberg~\citep{varshney-steinberg-2019} 
measured the correlation function
$G_{xy}(r,t)$ (as defined in equation~\ref{eq:Cxx}) in their experiments, and 
showed that the peaks of this function follow a linear scaling. 
The same behavior is observed in figure~\ref{fig:correla}h -- 
consistent with the experimental results. 
Yet, this scaling only applies to the peaks, not
to the whole correlation function (see the absence of collapse in the inset of 
figure~\ref{fig:correla}g)
or to other correlation functions, e.g., $G_{xx}$.
Furthermore, Alfv\'en waves require the
nonlinear advection term to exist, but we have shown that simulations without 
this term provide the same scaling exponent $\alpha = 4$.
Thus, these two points together show that elastic waves
{with linear dispersion relation} do not play
a role in determining the scaling behavior of ET.

Finally, in this paper, we clear the confusion of many different scaling
exponents for the energy spectrum of ET.
As shown in \Fig{fig:exponents}, over the years experiments and
simulations have obtained a large range of scaling exponents.
With time, the spatial exponent $\alpha$ is converging to $4$,
but the temporal exponent $\beta$ varies over a large range.
We now establish that, due to breakdown of Taylor's hypothesis,
$\beta$ is not universal. 
The focus of theory in ET should be to understand the $\alpha \approx 4$
scaling of the spatial energy spectrum.

\section*{Acknowledgments}
The research was supported by the Okinawa Institute of Science and Technology Graduate University (OIST) with subsidy funding to M.E.R. from the Cabinet Office, Government of Japan. M.E.R. also acknowledges funding from the Japan Society for the Promotion of Science (JSPS), grants 24K00810 and 24K17210. The authors acknowledge the computer time provided by the Scientific Computing Section and Data Analysis  section of the Core Facilities at OIST, and the computational resources provided by RIKEN, under the HPCI System Research Project Grants hp250021, hp250035, hp230018, hp220099, hp210269, and hp210229. DM acknowledges the support of the Swedish Research Council Grant No. 638-2013-9243. DM gratefully acknowledges hospitality by OIST. M.E.R. and G.F.R. acknowledge the useful discussion with Prof. Guido Boffetta and Stefano Musacchio.

\begin{figure*}[htb]
\centering
\includegraphics[width=.75\textwidth]{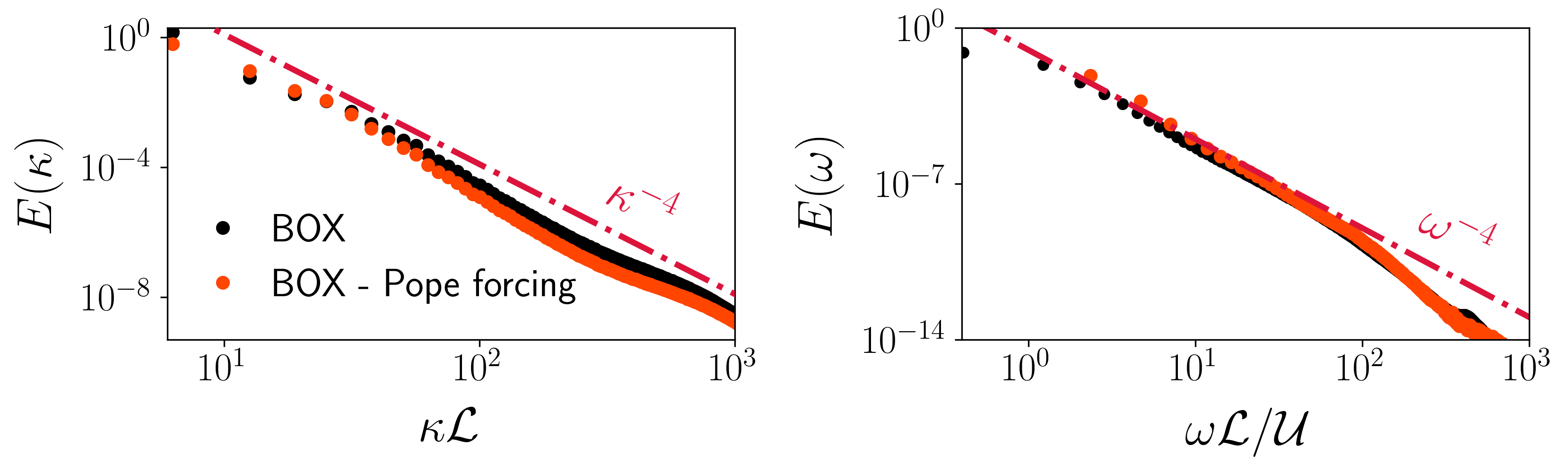}
\caption{\textbf{Turbulent kinetic energy spectra with Pope's forcing} in our box setup, normalised with the mean-square of the velocity fluctuations.}
\label{fig:Pope}
\end{figure*}

\section*{Author contributions}
G.F.R. analysed the data. G.F.R., R.K.S., D.M., and M.E.R. developed the theory, with inputs from A.C.. G.F.R., C.A., G.S., and M.E.R. developed the simulation code. G.S. and M.E.R. performed the numerical simulations. G.F.R., D.M., and M.E.R. outlined the manuscript content and finalised the manuscript, with the first draft written by G.F.R., and inputs from all authors. D.M. and M.E.R. conceived the original idea and supervised the research.

\appendix

\section{Independence of the results from the forcing}\label{app:pope}

To confirm the independence of our results from the forcing adopted, we replicated them with the one proposed by Eswaran and Pope \cite{eswaran-pope-1988}, observing no significant variation in the spatial and temporal spectra of the turbulent kinetic energy (figure~\ref{fig:Pope}).
No significant variation is observed in the spectral decay, regardless of the forcing adopted (ABC in black, Pope in orange), thus confirming the generality of our observations. Both the spatial (left) and temporal (right) spectra approach the power-law decay with exponent $-4$ discussed in the main text.
Note that this might not hold in two dimensions, where $\sexp$ was shown to vary between $-3$ and $-4$ depending on the forcing chosen \cite{plan-etal-2017,gupta-vincenzi-2019,yerasi-etal-2024}.

\section{Theory for the dynamics of ET in a periodic box}\label{app:theory}

We consider the Oldroyd-B equation~\citep{bird-1987} in a homogeneous and
isotropic setup, forced by an external stirring:
\begin{subequations}
  \begin{align}
    \rhof\left( \delt \ua + \ub \delb \ua \right) &= \notag\\
    -\dela p +
    \delb \left( 2\muf\Sab  + \frac{\mup}{\taup} \Cab\right) +\rhof\fa
     \label{eq:NN1}  \\ 
    \delt\Cab + \ug\delg\Cab &= \notag\\
    C_{\alpha\gamma}\delg \ub + C_{\gamma\beta}\dela\ug
    - \frac{1}{\taup}\left(\Cab - \dab\right)  \/.\label{Conf1} 
  \end{align}
  \label{eq:oldB}
\end{subequations}
Here, $\uu$ is the incompressible solvent velocity field, i.e.
$\delb \ub = 0$,
$p$ is the pressure,
$\mS$ is the rate-of-strain tensor with components
$\Sab \equiv (\dela\ub + \delb\ua)$,
$\muf$ is the dynamic viscosity of the solvent and
$\nuf = \muf/\rhof$ is its kinematic viscosity,
$\mup$ is the polymer viscosity,
$\rhof$ is the density of the solvent fluid,
$\taup$ is the polymer relaxation time, and
$\mathbf{C}$ is the polymer conformation tensor whose trace $\Cgg$
is the total end-to-end squared length of the polymer.
We use Einstein's notation of repeated indices being summed
and the symbol $\avg{\cdot}$ denotes averaging over turbulence. 

\subsection{Non-dimensionalization}
We use the following scheme for non-dimensionalization:
\begin{equation}
  \xx^{\ast} = \frac{\xx}{\mL}\/,\quad
  t^{\ast}=\frac{t}{\taup}\/,\quad
  \uu^{\ast} = \frac{\uu}{\mU}\/, \quad
  p^{\ast} = \frac{p}{\rhof \mU^2}\/.
  \label{eq:nondim}
\end{equation}
We denote by the superscript $\ast$ the non-dimensional form of
any quantity.
The tensor $\Cab$ is already dimensionless. 
For the homogeneous and isotropic case we use $\mU=\urms$
and non-dimensionalize the external force by its amplitude $F$.
Consequently the non-dimensional form of the PDEs (equation~\ref{eq:oldB}) becomes:
\begin{subequations}
  \begin{align}
    \frac{1}{\Deb}\delt^{\ast} \ua^{\ast}
    + \ub^{\ast} \delb^{\ast} \ua^{\ast}  &= \notag\\
    -\dela^{\ast} p^{\ast} +
    \frac{\mB}{\Rey}\delb^{\ast}2\Sab^{\ast}
    + \frac{1-\mB}{\Rey\Deb}\delb^{\ast}\Cab^{\ast} 
    + \frac{1}{\Fr}\fa^{\ast},
    \label{eq:NN2}  \\ 
    \frac{1}{\Deb}\left[\delt^{\ast}\Cab^{\ast} +\Cab^{\ast} - \dab \right]&= \notag\\
    -\ug^{\ast}\delg^{\ast}\Cab^{\ast} +
    C^{\ast}_{\alpha\gamma}\delg^{\ast} \ub^{\ast} +
    C^{\ast}_{\gamma\beta}\dela^{\ast}\ug^{\ast}\/.
    \label{Conf2} 
  \end{align}
  \label{eq:oldBndim}
\end{subequations}
Four dimensionless numbers govern the qualitative behavior of these
equations:
the Reynolds number $\Rey = \mU \mL/\nuf$;
the Deborah number $\Deb = \mU\taup/\mL$; 
the solvent-to-total-viscosity ratio $\mB = \muf/(\mup+\muf)$;
and the Froude number $\Fr = \mU^2/(\mL F)$. 
As we simulate a dilute polymer solution, we choose a fixed 
value for $\mB$ in all our simulations.

\subsection{Small $\Rey$ approximation}

To study ET we choose the Reynolds number to be small and the Deborah
number to be sufficiently large. 
We hold $\Deb$ to be constant, take the limit $\Rey\to 0$ and substitute 
$p^{\dagger} = \Rey\enspace p^{\ast}$~\citep{childress1981mechanics}.
In taking this limit, we assume that $\Rey/\Fr$ remains unity.
We then obtain 
\begin{subequations}
  \begin{align}
    \mB \left(\lap\right)^{\ast}\ua^{\ast} &= \notag\\
    \dela^{\ast} p^{\dagger} 
    - \frac{1-\mB}{\Deb}\delb^{\ast}\Cab^{\ast} - \fa^{\ast}, 
    \label{eq:NN3}  \\ 
    \frac{1}{\Deb}\left[\delt^{\ast}\Cab^{\ast} +\Cab^{\ast} - \dab \right]&= \notag\\
    -\ug^{\ast}\delg^{\ast}\Cab^{\ast} +
    C^{\ast}_{\alpha\gamma}\delg^{\ast} \ub^{\ast} +
    C^{\ast}_{\gamma\beta}\dela^{\ast}\ug^{\ast}\/.
    \label{Conf3} 
  \end{align}
  \label{eq:oldBndim2}
\end{subequations}
Here, we have assumed incompressibility to simplify the viscous stress
and the external force. 
From here on, we drop the superscripts -- it is assumed that the variables
are dimensionless. The equations are thus rewritten as:
\begin{subequations}
  \begin{align}
    \dela\ua = 0, \\
    \mB \lap\ua = \dela p 
    - \frac{1-\mB}{\Deb}\delb\Cab + \fa,
    \label{eq:NN4}  \\ 
    \delt\Cab = \dab-\Cab \notag\\
    +\Deb\left[C_{\alpha\gamma}\delg \ub + 
    C_{\gamma\beta}\dela\ug -\ug\delg\Cab \right].
    \label{eq:Conf3} 
  \end{align}
  \label{eq:oldBndim3}
\end{subequations}
In the absence of polymers, $\mup =0$, $\mB = 1$. Consequently,
equation~\ref{eq:oldBndim3} reduces to the incompressible, forced Stokes equation.

\subsection{In Fourier space}
Next, we Fourier transform in space (but not in time) and impose
incompressibility by multiplying equation~\ref{eq:NN4} by the projection operator
\begin{equation}
  \mP_{\ab}(\kk) \equiv \dab - \frac{\ka\kb}{k^2}\/,
  \label{eq:proj}
  \end{equation} where $\kk$ is the Fourier vector. 
We obtain:
\begin{equation}
  \mB k^2\ub(\kk)  = -\mathrm{i}\frac{1-\mB}{\Deb}\mP_{\ab}\kg C_{\alpha\gamma} + \mP_{\ab} \fa,
  \label{eq:NN5}
\end{equation}
We use the same symbol for variables and their Fourier transforms. 
The Fourier transform of equation~\ref{eq:Conf3} is
\begin{equation}
\begin{split}
  \delt\Cab(\kk) = \dab\delta(\kk)-\Cab(\kk) \\
  -\mathrm{i} \Deb\int\dpdq \, \delta(\kk-\pp-\qq)\\
  \left[\pg\ub(\pp)C_{\alpha\gamma}(\qq) +
      \pa\ug(\pp)C_{\gamma\beta}(\qq) -\qg\ug(\pp)\Cab(\qq) \right]\/.
\label{eq:Conf4}
\end{split}
\end{equation}    
Substituting equation~\ref{eq:NN5} to equation~\ref{eq:Conf4} we obtain
\begin{subequations}
\begin{align}
\begin{split}
\delt\Cab(\kk) = \dab\delta(\kk) - \Cab(\kk) \\
\quad - \frac{1-\mB}{\mB}\Ipq
\Bigg[
  \frac{\pk\pg}{p^2}\mP_{\delta\beta}C_{\delta\gamma}(\pp)C_{\alpha\kappa}(\qq) \\
+ \frac{\pa\pg}{p^2}\mP_{\delta\kappa}C_{\delta\gamma}(\pp)C_{\kappa\beta}(\qq) \\
\quad - \frac{\qg\pk}{p^2}\mP_{\delta\gamma}C_{\delta\kappa}(\pp)\Cab(\qq) 
\Bigg] \\
\quad - \mathrm{i} \frac{\Deb}{\mB}\Ipq \frac{1}{p^2} 
\Big[
  \pg\fb(\pp)C_{\alpha\gamma}(\qq) 
  + \pa\fg(\pp)C_{\gamma\beta}(\qq) \\
\quad - \qg\fg(\pp)\Cab(\qq)
\Big] ,
\end{split}
\label{eq:Conf5}
\end{align}
\end{subequations}
where $\quad \Ipq \equiv \int \dpdq\delta(\kk-\pp-\qq)\/$,    
and the argument of $\mP$,  $\pp$, is not explicitly written.
Here $d=3$ is the dimension of space. 
Altogether this is an intriguing term, which demonstrates the following.
One, even when the external force is limited to a single Fourier shell,
it appears in a convolution in equation~\ref{eq:Conf5}, i.e., its effects are felt
at all scales.
Two, the force multiplies the tensor $\Cab$, hence if the force is
random and Gaussian it appears as a \textit{multiplicative noise}. 

\begin{figure*}[htb]
\centering
\includegraphics[width=.75\textwidth]{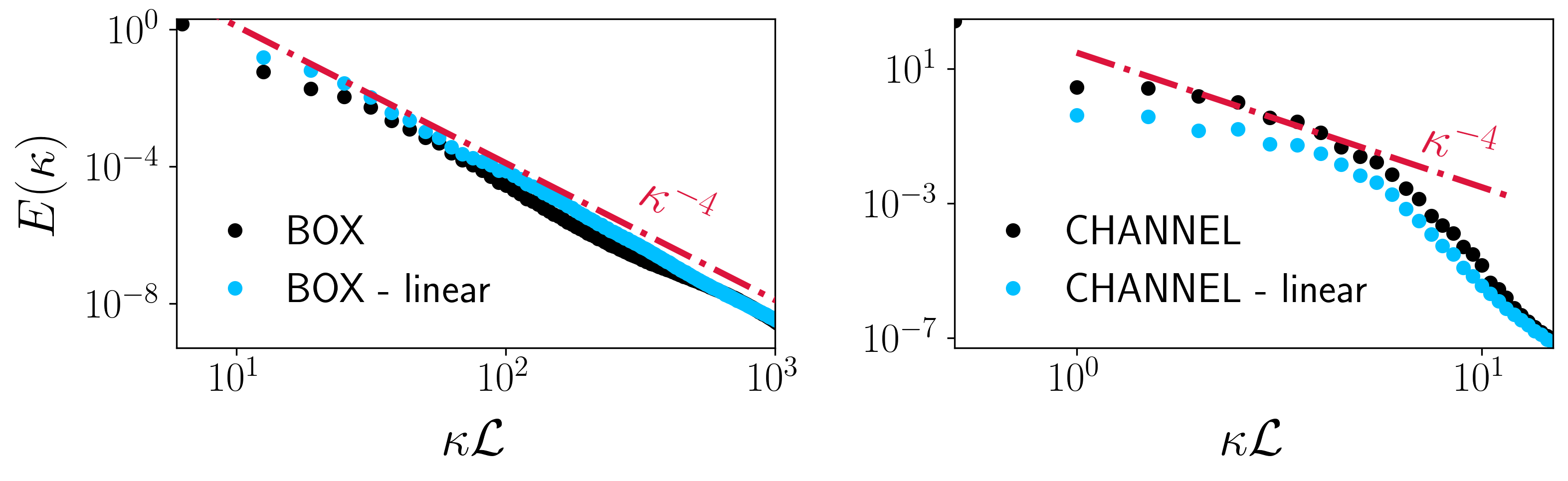}
\caption{\textbf{Turbulent energy spectra with and without nonlinear 
	advection}  
in box (left) and channel (right) setups. 
In black: spectra from solving the Navier-Stokes equation 
together with the Oldroyd-B model, equation~\ref{eq:oldB}. 
In blue: spectra with the nonlinear advection term ignored.
The two spectra have close overlap, both show $k^{-4}$ scaling.  
The spectra are normalised with the mean-square of the velocity fluctuations.} 
\label{fig:linear}
\end{figure*}

\subsection{Scaling analysis}
Let us now perform a straightforward scaling
analysis (see e.g., \citep[][section 9.6.4]{frisch-1995})
of equation~\ref{eq:Conf5}.
We rewrite the equations as
\begin{subequations}
\begin{align}
\begin{split}
\delt\Cab(\kk) = \dab\delta(\kk) - \Cab(\kk) \\ 
\quad + \boxempty(C,C) + \bigtriangleup(C,F),
\end{split} \\[1ex]
\begin{split}
\boxempty(C,C) = -\frac{1-\mB}{\mB}\Ipq
\Bigg[
  \frac{\pk\pg}{p^2}\mP_{\delta\beta}C_{\delta\gamma}(\pp)C_{\alpha\kappa}(\qq) \\
\quad + \frac{\pa\pg}{p^2}\mP_{\delta\kappa}C_{\delta\gamma}(\pp)C_{\kappa\beta}(\qq) \\ 
\quad - \frac{\qg\pk}{p^2}\mP_{\delta\gamma}C_{\delta\kappa}(\pp)\Cab(\qq)
\Bigg],
\end{split} \\[1ex]
\begin{split}
\bigtriangleup(C,F) = -\mathrm{i} \frac{\Deb}{\mB}\Ipq \frac{1}{p^2}
\Big[
  \pg\fb(\pp)C_{\alpha\gamma}(\qq) \\ 
\quad + \pa\fg(\pp)C_{\gamma\beta}(\qq) \\
\quad - \qg\fg(\pp)\Cab(\qq)
\Big],
\end{split}
\label{eq:dcdt}
\end{align}
\end{subequations}
Now, consider the scaling
\begin{equation}
  x\to bx\/,\quad t\to b^z t\/,\quad\text{and}\quad k \to k/b,
  \label{eq:sc}
\end{equation}
so that the fields in real space scale as 
\begin{equation}
  \uu(\xx) \to b^{h}\uu(\xx), \quad\text{and}\quad \Cab(\xx) \to b^{m}\Cab(\xx)\/.
  \label{eq:uCsc}
\end{equation}
The Fourier transforms of the same fields, of course, scale differently:
\begin{equation}
  \uu(\kk) \to b^{\chi}\uu(\kk), \quad\text{and}\quad \Cab(\kk) \to b^{\xi}\Cab(\kk)\/,
  \label{eq:uChsc}
\end{equation}
where $\chi = h+d$ and $\xi = m+d$, follows from the definition of
Fourier transform. 
We use a force that is constant in time and limited to a few small
Fourier modes.
Thus, under rescaling, the force remains constant. 
Then the different terms in equation~\ref{eq:dcdt} scale as
\begin{equation}
\begin{split}
\delt C_{\ab} \to b^{\xi-z}\delt C_{\ab}\/,\quad C_{\ab} \to b^{\xi}C_{\ab},\\
  \quad \boxempty \to b^{2\xi-d}\boxempty,
  \quad \bigtriangleup \to b^{\xi-d+1}\bigtriangleup.
  \label{eq:scdc}
\end{split}
\end{equation}
Here, we have ignored the term $\delta(\kk)$ as it is zero for all $\kk$
except for $\kk = 0$.
Assuming that equation~\ref{eq:dcdt} is scale invariant, we now obtain the
following relations
\begin{subequations}
  \begin{align}
    \xi - z = \xi &\implies z = 0, \\
    \xi -z = 2\xi -d  &\implies z = d-\xi, \\
    \xi -z = \xi-d+1 &\implies z = d-1 \/.
  \end{align}
\end{subequations}
From \eq{eq:NN4} we obtain: $h-2 = m -1 = \xi -d -1$, and 
substituting this in \eq{eq:NN4} and simplifying, we obtain the following three
choices:
\begin{equation}
  z = 0 \/,\quad z = 1-h \/,\quad\text{and}\quad z = d-1 = 2\/.
  \label{eq:zs}
  \end{equation}
In ET we know that $h=1$, hence the first two expressions are consistent
with each other, but the last one is not.
Even if the force is limited to a few Fourier modes, since the
term $\bigtriangleup$ is a convolution, its effects are present
in all modes and cannot be considered small anywhere.
Thus we conclude that there is no unique dynamic scaling exponent
in ET. 

\section{Neglecting the non-linear advection}\label{app:nlAdv}

To assess the role played in ET by the non-linear advection term,
$\ub\delb\ua$, we perform two additional simulations in the box and channel
setups where this term is explicitly removed from the momentum
equation.
We do not observe any significant change to the spatial spectra
of the turbulent kinetic energy in either of these cases,
see figure~\ref{fig:linear}. 
The spectra exhibit a clear $k^{-4}$ power-law decay
for the box (over almost two decades) and the channel (little more
than half a decade).

\bibliographystyle{unsrt}
\bibliography{./bibliography.bib}

\end{document}